\documentclass{appolb}
\usepackage{graphicx}
\usepackage{amsmath}
\usepackage{amssymb}
\usepackage{cite}
\AtBeginDocument{\providecommand\figref[1]{\ref{fig:#1}}}
\AtBeginDocument{}

\makeatother

\begin{document}
\title{
Computing the topological susceptibility from fixed topology QCD simulations%
\thanks{Presented at Excited QCD 2016}%
}
\author{Arthur Dromard$^{(1)}$, Wolfgang Bietenholz$^{(2)}$,\\ Krzysztof Cichy$^{(1),(3)}$, Marc Wagner$^{(1)}$
\address{$^{(1)}$~Goethe-Universit\"at Frankfurt am Main \\
Institut f\"ur Theoretische Physik \\
Max-von-Laue-Stra{\ss}e 1, D-60438 Frankfurt am Main, Germany}
\address{$^{(2)}$~Instituto de Ciencias Nucleares \\
Universidad Nacional Aut\'{o}noma de M\'{e}xico \\
A.P.\ 70-543, C.P.\ 04510 Ciudad de Mexico, Mexico}
\address{$^{(3)}$~Adam Mickiewicz University\\
Faculty of Physics\\
Umultowska 85, 61-614 Poznan, Poland}
}
\maketitle
\begin{abstract}
The topological susceptibility is an important quantity in QCD,
which can be computed using lattice methods. However, at a fine lattice
spacing, or when using high quality chirally symmetric quarks, algorithms which proceed in
small update steps --- in particular the HMC algorithm ---
tend to get stuck in a single topological
sector. In such cases, the computation of the topological susceptibility
is not straightforward. Here, we explore two methods to extract
the topological susceptibility from lattice QCD simulations restricted to
a single topological sector. The first method is based on the
correlation function of the topological charge density, while the second
method relies on measuring the topological charge within spacetime
subvolumes. Numerical results for two-flavor QCD obtained by using both methods are
presented.
\end{abstract}


\section{Introduction}

In lattice QCD simulations with periodic boundary conditions, the autocorrelation time of the topological charge can be rather large. In particular, this is the case for lattice spacings
$a \lesssim 0.05 \, \textrm{fm}$, where the topological charge is typically frozen \cite{Luscher:2011kk}.
For chirally symmetric quark actions, such a freezing takes place also at much coarser lattice spacings (cf.\ e.g.\ \cite{Aoki:2012pma}).
To overcome this problem, simulations with open boundary conditions have been advocated \cite{Luscher:2011kk}. However, even though promising, they might not always be applicable. 
For example, when using a mixed action setup with light overlap valence
and corresponding Wilson sea quarks, it is extremely difficult to take the continuum limit correctly, since exact zero modes of the valence Dirac operator are not compensated by the sea quark determinant \cite{Cichy:2010ta}. A possible solution to this problem is to use topology conserving
actions (cf.\ e.g.\ \cite{Fukaya:2005,Bietenholz:2005rd,Bruckmann:2009cv}) and to simulate only the topological sector $Q=0$, where zero modes are absent (in stochastic configurations).

In this work, we use periodic boundary conditions and explore two methods to extract the topological
susceptibility from simulations confined to a single topological sector. The first
one is the Aoki-Fukaya-Hashimoto-Onogi (AFHO) method \cite{Aoki:2007ka}, which allows
for the extraction of the topological susceptibility from the correlation function of the topological
charge density. This method has already been studied in several models and theories including SU(2) Yang-Mills theory \cite{Bautista:2015yza}. The second approach is the slab method, where the topological susceptibility is determined from computations of the topological charge on spacetime subvolumes. The  method was sketched in \cite{deForcrand:1998ng} and has recently been tested in lower dimensional models \cite{Bietenholz:2015rsa}. We will present numerical tests for two-flavor QCD using both methods.

\section{\label{SEC001}Computation of $\chi_t$ at unfixed topology}

To be able to verify that the fixed topology methods yield correct results, we have
first computed the topological susceptibility at unfixed topology.

To this end we have generated 10000 gauge link configurations using two-flavor Wilson twisted mass lattice QCD (cf.\ \cite{Jansen:2009xp} for details regarding the simulation code). The lattice spacing is $a \approx 0.079 \, \textrm{fm}$, the pion mass is $m_\pi \approx 650 \, \textrm{MeV}$ and the lattice volume is $16^3 \times 32$.

On each gauge link configuration, the topological charge has been computed
using the field strength definition (cf.\ e.g.\ \cite{Bietenholz:2016ymo}), but with the gradient flow instead of cooling to reduce UV-fluctuations. Moreover, the topological charge has been renormalized by locally minimizing $\langle(\alpha Q_L-\textrm{int}(\alpha Q_L))^{2}\rangle$ with respect to $\alpha$, in the region  $\alpha\approx 1$ ($\alpha$ is a multiplicative renormalization parameter), $Q_L$ is the measured topological charge and $\textrm{int}(x)$ denotes the integer closest to $x$.

The topological susceptibility has then been computed via
\begin{equation}
\chi_t = \frac{\langle Q^2\rangle}{V},
\end{equation}
where $V$ is the spacetime volume and the statistical error has been determined by a bootstrap analysis. Additionally, a systematic error has been estimated by comparing $\chi_t$ for different flow times. Our result is $\chi_t a^4= 7.76(20) \times 10^{-5}$ (since we are mainly interested in testing numerical methods, we always quote $\chi_t$ in units of the lattice spacing).

\section{Computation of $\chi_t$ with the AFHO method}

The AFHO method is based on the topological charge density correlation function at large separations computed in a fixed topological sector, 
\begin{equation}
\label{EQN001}\Big\langle q(t) q(0) \Big\rangle_{|Q|,V} \underset{t \rightarrow \infty}{=} -\frac{\chi_{t}}{V}\left(1-\frac{Q^{2}}{\chi_{t}V}\right)+\mathcal{O}\left(\frac{1}{\chi_{t}^{2}V^{2}},e^{-m_{\eta}t}\right) ,
\end{equation}
where $q$ is the topological charge density and $m_\eta$ the mass of the lightest $I(J^P) = 0(0^-)$ meson. Eq.\ (\ref{EQN001}) is an expansion in $1 / (\chi_t V)$ and $Q^2 / (\chi_t V)$, i.e.\ is valid for $1 / (\chi_t V) \ll 1$ and $Q^2 / (\chi_t V) \ll 1$. From the unfixed
topology result of Section~\ref{SEC001} follows $\chi_t V \approx 10$, which implies that (\ref{EQN001}) should be quite accurate for $|Q|<3$, while $|Q|=3$ has to be treated with caution and $|Q|>3$ must be discarded. Eq.\ (\ref{EQN001}) shows that at large separations, the topological charge density correlation function converges to a constant, from which one can easily extract $\chi_t$.

\begin{figure}[b]
\includegraphics[angle=270,width=12.5cm, totalheight=8cm]{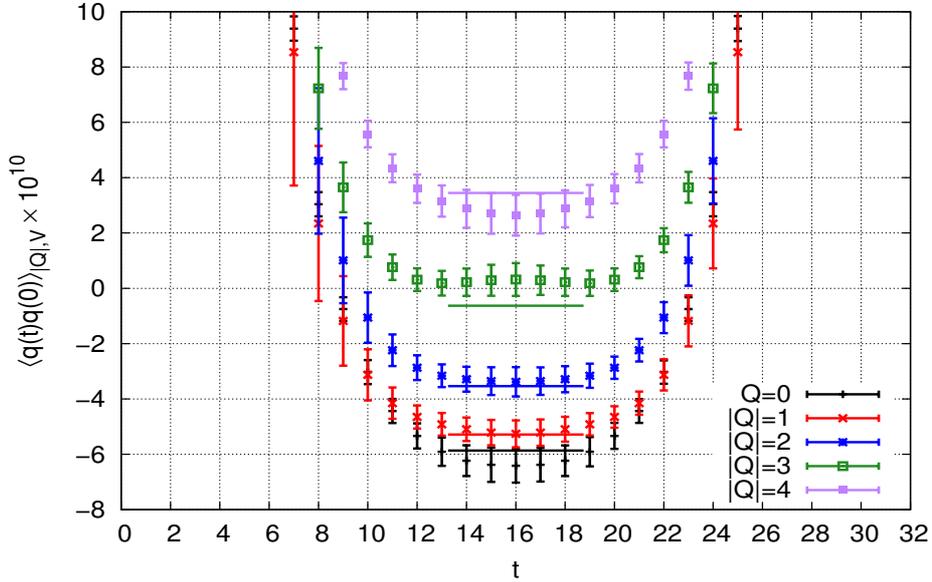}

\caption{\label{fig:AFHO}$\langle q(t)q(0)\rangle_{|Q|,V}$ as a function of
the separation $t$ at flow time $\tau = 6 \tau_0$ for different values of
the topological charge $Q$. The horizontal lines are the expected plateaus $-(\chi_t / V) (1 - Q^2 / (\chi_t V))$ with $\chi_ta^4 = 7.76 \times 10^{-5}$ from Section~\ref{SEC001}. }
\end{figure}

Our results have been obtained using the same gauge link configurations as in Section~\ref{SEC001}. In Fig.~\figref{AFHO}, our numerical results are shown, the topological charge density correlator $\langle q(t)q(0) \rangle_{|Q|,V}$ as a function of the separation $t$ for different topological sectors. 
The theoretically expected values, using $\chi_t a^4= 7.76(20) \times 10^{-5}$ from the previous section, have also been plotted and are represented by the horizontal lines. The plateau values corresponding to different $Q$, as predicted
in eq.~(\ref{EQN001}), are clearly distinct. For values with $Q^2 / (\chi_t V) \ll 1$ (i.e.\ $|Q| \leq 2$), there is excellent agreement. On the other hand, for $|Q| \geq 3$ there is a slight tension, probably a consequence of $Q^2 / (\chi_ t V) \geq 0.88$, which is the expansion parameter of eq.\ (\ref{EQN001}) and which should be small. We obtain $\chi_t a^4 = 7.69(22) \times 10^{-5}$ via a combined fit to the $Q=0$, $|Q|=1$ and $|Q|=2$ results, which is consistent with our unfixed topology result from Section~\ref{SEC001}, $\chi_t a^4= 7.76(20) \times 10^{-5}$.

\section{Computation of $\chi_t$ with the slab method}

Assuming a Gaussian distribution of the topological charge, this method uses
spacetime subvolumes $x V$, $x \in [0,1]$ called ``slabs''. The probability of having
topological charge $\bar Q \in \mathbb{R}$ inside a slab under the condition that the topological charge of the total volume is $Q$ is straightforward to calculate, 
\begin{equation}
\label{EQN010} p(\bar{Q})p(Q-\bar{Q})\bigg|_{xV,Q}\propto\exp\bigg(-\frac{\bar{Q}^{2}}{2\chi_{t}Vx}\bigg)\times\exp\bigg(-\frac{(Q-\bar{Q})^{2}}{2\chi_{t}V(1-x)}\bigg).
\end{equation}
Defining $\bar{Q}'=\bar{Q}-xQ$ allows to simplify (\ref{EQN010}),
\begin{equation}
p(\bar{Q})p(Q-\bar{Q})\bigg|_{xV,Q}\propto\exp\bigg(-\frac{\bar{Q}'^{2}}{2\chi_{t}Vx(1-x)}\bigg).
\end{equation}
From this expression one can read off
\begin{equation}
\label{EQN002}\Big\langle \bar{Q}'^{2} \Big\rangle=\chi_{t}Vx(1-x).
\end{equation}
The method to extract the topological susceptibility is then straightforward:
one has to compute $\langle \bar{Q}'^{2}\rangle$, the average of $\bar{Q}'^{2}$
on the available gauge link configurations with topological charge $Q$, for several values of
$x$. The resulting points should be consistent with the parabola (\ref{EQN002}), i.e.\ $\chi_t$ can be obtained with a corresponding fit.

\begin{figure}[t]
\includegraphics[angle=270,width=12.5cm, totalheight=8cm]{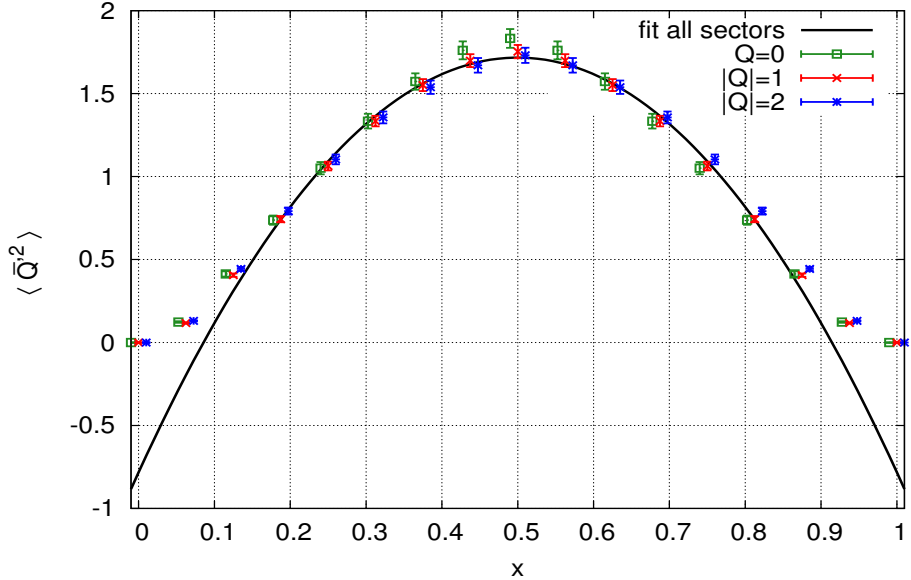}
\caption{\label{fig:Slab}$\langle \bar{Q}'^{2} \rangle$ as a function of $x$ for different values of the topological charge $Q$ (for better visibility points for $Q=0$ ($|Q|=2$) are slightly shifted to the
left (right)). The black curve represents the fit of eq.\ (\ref{EQN002}) with an additive constant to the data points.}
\end{figure}

In Fig.~\figref{Slab}, numerical results for $\langle \bar{Q}'^2 \rangle$ are shown for $|Q| = 0,1,2$.
The slabs used for the computations have temporal extent $x T$ and spatial volume $L^3$, periodic in space. We observe that the data points are not fully consistent with the quadratic curve (\ref{EQN002}). In particular at small $x$ and small $1-x$ there are strong discrepancies. On the other hand, the data points in the interval $0.2 \leq x \leq 0.8$ can be described nicely with (\ref{EQN002}), if one allows for an additive constant. Moreover, the corresponding result, $\chi_t a^4 = 7.63(14) \times 10^{-5}$, is then in agreement with the unfixed topology result from Section~\ref{SEC001}, $\chi_t a^4= 7.76(20) \times 10^{-5}$. The distortions at small $x$ and small $1-x$ seem to be related to the gradient flow: for large flow times, these distortions become more and more prominent. Understanding this issue in detail is part of our current research.

\section{Summary and conclusions}

To conclude, we have presented two methods allowing for the extraction of the topological
susceptibility from simulations within a single topological sector. We have successfully applied
them to QCD, demonstrating their practical use. The slab method has 
the advantage that one can apply it to any sufficiently large volume without encountering additional problems, while for the AFHO method the signal-to-noise ratio becomes worse when increasing the volume (cf.\ \cite{Bautista:2015yza} for a detailed discussion).

\section*{Acknowledgments}

A.D.\ and M.W.\ acknowledge support by the Emmy Noether Programme of the DFG (German Research Foundation), grant WA 3000/1-1. 
W.B.\ acknowledges support by DGAPA-UNAM, grant IN107915.
K.C.\ was supported in part by the Deutsche Forschungsgemeinschaft (DFG), project nr. CI 236/1-1 (Sachbeihilfe).
 
This work was supported in part by the Helmholtz International Center for FAIR within the framework of the LOEWE program launched by the State of Hesse. Calculations on the LOEWE-CSC high-performance computer of Johann Wolfgang Goethe-University Frankfurt am Main were conducted for this research. We would like to thank HPC-Hessen, funded by the State Ministry of Higher Education, Research and the Arts, for programming advice.


 
 \end{document}